\documentclass[preprint]{elsarticle}
\usepackage{amsmath}
\usepackage{graphicx}
\usepackage{hyperref}
\usepackage{amssymb}
\usepackage{mathtools}
\usepackage{mathrsfs}
\usepackage{amsfonts}
\usepackage{bbm}
\usepackage{enumerate}
\usepackage{array}
\usepackage{booktabs}
\usepackage{textcomp}
\usepackage{caption}
\usepackage{cancel}
\usepackage{units}
\usepackage{multirow}
\usepackage{pst-all}
\usepackage{longtable}
\usepackage{relsize}
\usepackage{}[natbib]

\begin{document}
\begin{frontmatter}
\title{Partition Function of N=2  Gauge Theories on an Squashed $S^4$
with $SU(2)\times U(1)$ Isometry}

\author[address1]{Alejandro Cabo-Bizet\fnref{f}}
\fntext[f]{acabo@sissa.it}
%\fntext[$\P$]{acabo@sissa.it}
\author[address2]{Edi Gava\fnref{e}}
\fntext[e]{gava@sissa.it}
\author[address1,address2,address3]{Victor I. Giraldo-Rivera\fnref{g}}
\fntext[g]{vgiraldo@ictp.it}
\author[address3]{M. Nouman Muteeb\fnref{h}}
\fntext[h]{mmuteeb@sissa.it}
\author[address1]{K.S. Narain\fnref{i}}
\fntext[i]{narain@ictp.it}
\address[address1]{ICTP, Strada Costiera 11, 34014 Trieste, Italy}
\address[address2]{INFN, sezione di Trieste, Italy}
\address[address3]{SISSA,Via Bonomea 265, 34128 Trieste, Italy}

\begin{abstract}
 We study $N=2$ supersymmetric gauge theories on a large
  family of squashed 4-spheres preserving $SU(2)\times U(1)\subset
  SO(4)$ isometry and determine the conditions under which this
  background is supersymmetric. We then compute the partition function
  of the theories by using localization technique. The results
  indicate that for  $N=2$ SUSY, including both vector-multiplets and
  hypermultiplets,  the partition function is independent of the
  arbitrary  squashing functions as well as of the other supergravity
  background fields. 
\end{abstract}

\end{frontmatter}

\section{Introduction}
Supersymmetric Localization techniques furnish a rich ground for exact computation of various quantities in Supersymmetric Quantum Field Theories. This program started with the work
of \cite{Witten:1992xu}, later pursued by
\cite{Nekrasov:2003af,Nekrasov:2003rj} and more recently brought back by
\cite{Pestun:2007rz}  which gave rise to an intense activity of exact
calculations in various dimensions and/or manifolds.
\cite{Kapustin:2009kz,Yaakov:2012usa,Gupta:2012cy,Hama:2010av,Nosaka:2013cpa,Gomis:2012wy,Hama:2012bg,Imamura:2012bm}.
A systematic way to put rigid SUSY on curved spaces in the case of
$N=1$ theories  was worked out by
\cite{Festuccia:2011ws,Dumitrescu:2012ha}, and, for
$N=2$ theories in \cite{Klare:2013dka}. 
%this  complemented the
%extensive program  to calculate several observables as partition
%functions, expectation values of Wilson loops  and other non-local
%operators. 
The partition function on squashed spheres depend in
general on the squashing parameters \cite{Hama:2010av,Hama:2012bg}.
However for some squashing, preserving a particular isometry of the
manifold,  the partition function comes out to be independent of
squashing parameters. Detailed studies of 3-dimensional cases had
appeared \cite{Hama:2010av}, \cite{Alday:2013lba},
\cite{Closset:2013vra} and
\cite{Nishioka:2013haa, Huang:2014gca, Imamura:2013nra}.
%For example \cite{Hama:2010av} considered
%squashed $S^3$ with $SU(2)\times U(1)$ isometry and showed that the
%$N=2$ supersymmetric partition function was independent of the
%squashing. There the  squashing independence of the partition function was due to the fact \cite{Hama:2010av} that the eigenmodes of the  kinetic operators for scalars and fermions form
%multiplets of same eigenvalues under $SU(2)$ isometry group. A deeper
%understanding of this fact came with \cite{Alday:2013lba} and then \cite{Closset:2013vra}, in the former the partition function of $N=2$ theories on three manifolds, was shown to depend on the Reeb vector field associated to the contact structure,  while in the latter the partition function depends on an equivalent geometrical structure called transversely
%holomorphic foliation. The $SU(2)\times U(1)$ squashing of $S^3$ does
%not change these  geometrical structures, while the
%$U(1)\times U(1)$ squashing 
%does it, and hence in the   $SU(2)\times U(1)$ squashed-partition function  there is no squashing parameter dependence, while in $U(1)\times U(1)$- partition function  there is. Other studies on squashed $S^3$ have been
%performed  in \cite{Nishioka:2013haa,Huang:2014gca,Imamura:2013nra},
%with applications to SUSY R\`{e}nyi entropy and holography.\\
For the four dimensional case, the analysis of which  geometrical background  data
 the partition function depends on,  has
 been performed for $N=1$ SUSY. 
The four dimensional squashed sphere has also been considered, first in
\cite{Hama:2012bg}, and later in
\cite{Nosaka:2013cpa,Huang:2014pda,Crossley:2014oea}.
The SUSY partition function on the branched $S^4$ in \cite{Nosaka:2013cpa,Huang:2014pda} computes the SUSY  R\`{e}nyi entropy of a circular region in a 4-dimensional space \cite{Casini:2011kv,Klebanov:2011uf}.\\
In this paper we calculate $N=2$ supersymmetric partition function on
a very general squashed $S^4$ with $SU(2)\times U(1)$ isometry, and
show that it is independent of the squashing metric parameters and of the
other supergravity backgrounds. In the case of $N=2$  theories on the ellipsoid considered in  \cite{Hama:2012bg} the isometry is  generically $U(1)\times U(1)$. In the limit $l=r$ in \cite{Hama:2012bg} this symmetry is enhanced to the $SO(3)\times SO(2)$ subgroup
of $SO(5)$. 
%which is to be seen as the subgroup of $SO(5)$. 
On the other hand, the $SU(2)\times U(1)$ isometry in our case is a subgroup of $SO(4)\equiv SU(2)_L\times SU(2)_R$.
The paper is organized as follows. In section \ref{susy curve}, the
Killing spinor equations for $N=2$ rigid SUSY on squahsed $S^4$ are
given, in section \ref{Squashed $S^4$} the squashed $S^4$ metric and
spin connection components are given and  we solve the  Killing spinor
equations, calculating various background fields and then  giving the conditions for their regularity. Section \ref{multiplets} contains the calculation of the $Q^2$ action on the fields of vector-multiplets and hypermultiplets. In sections \ref{path integration} and \ref{One-loop determinant}, we find
the saddle point configurations and one-loop determinants for
vector multiplet and hypermultiplet  respectively.
%, closely following
%\cite{Hama:2012bg} and \cite{Pestun:2007rz}. 
In section
\ref{instanton} we comment on the contribution of point-like  instantons and
anti-instantons to the supersymmetic  partition function. A brief summary of the main result is given in section  \ref{conclusion}.
\section{Rigid Supersymmetric Theories on Curved Spaces}\label{susy curve}
By now a systematic  way to put rigid SUSY on a curved spaces has been
developed: the procedure is  to start from  the  supergravity transformations  \cite{Festuccia:2011ws,Dumitrescu:2012ha,Dumitrescu:2012at,Klare:2013dka} and
obtain a rigid SUSY theory on a given  curved manifold  by 
freezing the quantum fluctuations of the gravitational background 
by taking the Planck mass limit $M_P\rightarrow \infty$, setting
to zero and the fermionic fields in the supergravity multiplet. 
We following this procedure for $N=2$
%of \cite{Hama:2012bg}, 
one obtaines a set of Killing spinor equations which have to be satisfied
in order to obtain rigid 4D $N=2$ SUSY  and at the same time constrain
the background geometry. They are:
%. We will consider another squashed-$S^4$, the formalism remains
%effectively the same as \cite{Hama:2012bg}.
%In \cite{Hama:2012bg} a couple of set of equations were obtained, the  Killing spinor and  auxiliary equations, the first by setting  the SUSY variation of the gravitino to zero
\begin{align}\label{eq:101}
\begin{split}
D_m\xi_A+T^{kl}\sigma_{kl}\sigma_m\bar{\xi}_A &=-\iota\sigma_m\bar{\xi}^\prime_A,\\ D_m\bar{\xi}_A+\bar{T}^{kl}{\bar{\sigma}}_{kl}{\bar{\sigma}}_m\xi_A &=-\iota {\bar{\sigma}}_m\xi^\prime_A\quad \text{for a given pair} \quad\xi^\prime_A,\bar{\xi}^\prime_A,
\end{split}\end{align}
(where $\iota \equiv \sqrt{-1}$)
coming from the gravitino variation, and:
\begin{align}\label{eq:103}
\begin{split}
\sigma^m{\bar{\sigma}}^n D_mD_n\xi_A+4D_lT_{mn}\sigma^{mn}\sigma^l\bar{\xi}_A &=M\xi_A,\\
{\bar{\sigma}}^m\sigma^n D_m D_n\bar{\xi}_A+4D_l {\bar{T}}_{mn}{\bar{\sigma}}^{mn}{\bar{\sigma}}^l\xi_A &=M\bar{\xi}_A,
\end{split}
\end{align}
with $M$  a real scalar background field, which is a consequence of the  variation of a spin 1/2
field in the supergravity multiplet.\\
Here $\xi_{A}$ and $\bar{\xi}_{A}$ (spinor indices are omitted) are chiral and anti-chiral Killing spinors satisfying
reality conditions to be specified later
%of  \cite{Hama:2012bg}, 
and are the parameters of
$N=2$ SUSY. The index $A$  is  a $SU(2)_R$
$R$-symmetry index of the $N=2$ theory. The fields $T^{kl}$, $\bar{T}^{kl}$ are a self-dual and anti-self-dual real tensor
background fields respectively. The covariant derivatives in \eqref{eq:101} and \eqref{eq:103} are covariantized also  with respect to a background $SU(2)_R$ gauge field $(V_m)^A_{\,{B}}$, in addition to the to the local Lorenz and gauge transformations.
We work in four component notation, where \eqref{eq:101} is written compactly as
\begin{align} 
D_m\xi+T.\Gamma \Gamma_m \xi=-\iota \Gamma_m \xi^\prime,
\end{align}
where $T.\Gamma\equiv T_{kl}\Gamma^{kl}$ \footnote{Our conventions of $\Gamma$ matrices can be simply read off the Killing spinor equations of \cite{Hama:2012bg} and their $\sigma$ matrices.}. Now multiplying from left by
$\Gamma^m$ and using the identity $\Gamma_m\Gamma_{kl}\Gamma^m=0$ we get
\begin{align}
\xi_p\equiv \Gamma^m D_m\xi=-4\iota \xi^\prime.
\end{align}
Here a new spinor $\xi_p$ is defined which will be useful later on,
when we will calculate the square of supersymmetry transformation $Q^2$ acting on different fields of $N=2$ theory.\\
\section{Supersymmetry on the Squashed $S^4$}\label{Squashed $S^4$}
The family of  squashed 4-spheres which we will consider are defined by the following
metric or vielbein one-forms:
\begin{align}
\begin{split}
&ds^2=\text{d$r$}^2+\frac{f(r)^2}{4} \left(\text{d$\theta $}^2+ \sin^2{\theta}\text{d$\phi $}^2 \right)+\frac{h(r)^2}{4} (\text{d$\psi $}+ \cos{\theta}\text{d$\phi $} )^2 ,\\
&e^4=\text{d$r$} , \quad  e^3=-\frac{h(r)}{2} \left(\text{d$\psi $}+\cos{\theta}\text{d$\phi $} \right),\quad  e^2=\frac{f(r)}{2} \left(\sin{\psi}\text{d$\theta $}- \sin{\theta}\cos{\psi} \text{d$\phi $} \right),\\ & e^1=-\frac{f(r)}{2} \left( \cos{\psi} \text{d$\theta $}+  \sin{\theta }\sin{\psi}\text{d$\phi $}\right),
\end{split}
\end{align}
% vielbein
where $f(r)$ and $h(r)$ are smooth arbitrary functions of $r$ . The above metric has $SU(2)\times U(1)$ isometry.
The  spin connection is given by the following non-zero components $\Omega _m^{\text{ab}}$,
\begin{align}
\begin{split}
&\Omega _1^{21}  =1-\frac{h(r)^2}{2 f(r)^2}, \quad \Omega _1^{43}=\frac{h'(r)}{2},\quad
\Omega _2^{31} =\frac{h(r) \sin (\psi )}{2 f(r)}, \quad \Omega _2^{32} = \frac{h(r) \cos (\psi )}{2 f(r)},\quad \\
& \Omega _2^{41} =\frac{1}{2} \cos (\psi ) f'(r),
\Omega _2^{42}=-\frac{1}{2} \sin (\psi ) f'(r),\quad \Omega _3^{21}=\cos (\theta )-\frac{h(r)^2 \cos (\theta )}{2 f(r)^2},\quad \\ & \Omega _3^{31}=-\frac{h(r) \sin (\theta ) \cos (\psi )}{2 f(r)}, \quad \Omega _3^{32}=\frac{h(r) \sin (\theta ) \sin (\psi )}{2 f(r)},\quad  \Omega _3^{41}=\frac{1}{2} \sin (\theta ) \sin (\psi ) f'(r),\quad \\ & \Omega _3^{42}=\frac{1}{2} \sin (\theta ) \cos (\psi ) f'(r),\quad \Omega _3^{43}=\frac{1}{2} \cos (\theta ) h'(r),
\end{split}
\end{align}
where $a,b=1,..,4$ are flat indices and $m=1,...4$ is curved space index.
\subsection{Solution of Killing Spinor Equation on the Squashed $S^4$}\label{killing solution}
%When a sphere is squashed, as in our case, the Killing spinor
%equation  needs in comparison to  the round sphere \cite{Pestun:2007rz} from additional background fields $T, V$ to cancel its torsion, the covariant derivative contains also the background gauge field $V$ \cite{Hama:2012bg}.
The purpose of this section is to show that if the background fields $(V_m)^A_{\,{B}}, T_{mn}, \bar{T}_{mn}, M $ are chosen appropriately, the squashed $S^4$ admits a Killing spinor which is solution of the two stets of Killing spinor equations \eqref{eq:101} and \eqref{eq:103}.
We write the backgrounds  $T$ and $ V$ in a complexified version:
% Background Fields V and T
\begin{align}
\begin{split}
&V_m =\left(
\begin{array}{cc}
 \iota v_{3m} & \iota (v_{1m}+\iota v_{2m}) \\
 \iota (v_{1m}-\iota v_{2m}) & -\iota v_{3m} \\
\end{array}
\right),\\
&T = \left(
\begin{array}{cccc}
 \iota t_3 & \iota (t_1-\iota t_2) & 0 & 0 \\
 \iota (t_1+\iota t_2) & -\iota t_3 & 0 & 0 \\
 0 & 0 & \iota t_3 & \iota (\bar{t}_1-\iota \bar{t}_2) \\
 0 & 0 & \iota (\bar{t}_1+\iota \bar{t}_2) & -\iota \bar{t}_3 \\
\end{array}
\right).
\end{split}
\end{align}
We will consider the following ans\"{a}tz for the Killing spinor and
we will calculate the background fields $T$, $V$ and $M$ such that this ans\"{a}tz
satisfies the set of Killing spinor equations
\begin{align}\label{eq:ansa}
\xi=\left(
\begin{array}{cc}
 s_1(r) & 0 \\
 0 & t_2(r) \\
 s_3(r) & 0 \\
 0 & t_4(r) \\
\end{array}
\right).
\end{align}
The Killing spinor satisfies the reality condition given in \cite{Hama:2012bg}:
\begin{align}\begin{split}
(\xi_{\alpha A})^{\dagger}=\epsilon^{AB}\epsilon^{\alpha\beta}\xi_{\beta B},\qquad
(\bar{\xi}_{\dot{\alpha} A})^{\dagger}=\epsilon^{AB}\epsilon^{\dot{\alpha}\dot{\beta}}\xi_{\dot{\beta} B}.
\end{split}\end{align}
The parameters in the Killing spinor are arbitrary smooth functions of $r$. After solving the Killing spinor equations, it turns out that
some of these parameters are constrained.\\
The general solution to the main and auxiliary
equations using the ans{\"a}tz \eqref{eq:ansa} takes
the following form: 
% Background fields
\begin{align}\begin{split}
&s_1(r)=s(r) ,\quad s_3(r)=\frac{\iota~ c~ h(r) }{s(r)},\quad t_2(r)=s(r) ,\\ &t_4(r)=-\frac{\iota~ c~ h(r) }{s(r)},\\
 & t_3 =\frac{s(r) \left(f(r) \left(2 f(r) s'(r)-s(r) f'(r)\right)+h(r) s(r)\right)}{4 c f(r)^2 h(r)},\\ & \bar{t}_3 =\frac{c \left(f(r) h(r) \left(s(r) f'(r)+2 f(r) s'(r)\right)-2 f(r)^2 s(r) h'(r)+h(r)^2 s(r)\right)}{4
   f(r)^2 s(r)^3},\\& v_{33} =\frac{1}{2} \left(\frac{h(r)}{f(r)^2}+\frac{h'(r)-2}{h(r)}-\frac{2 s'(r)}{s(r)}\right),\\ & M = \frac{2 f''(r)}{f(r)}+\frac{f'(r)^2-2 h'(r)+\frac{4 h(r) s'(r)}{s(r)}}{f(r)^2}+\frac{h(r)^2}{f(r)^4}+\frac{4 s'(r)
   \left(s(r) h'(r)-h(r) s'(r)\right)}{h(r) s(r)^2} .
\end{split}
\end{align}
Here only the non-zero part of the background fields and Killing spinor components
are given,  $c$ is a real  arbitrary constant which  sets
normalization of the killing vector we will to localize, $s(r)$ is a smooth function of $r$ and the background fields $T$
and $V_m$ are indexed by flat tangent space indices. For these
background fields to be well defined on the squashed $S^4$, it is
necessary that  $s(r)$ has no zero between the two poles.
We thus determined the form of all the additional background fields in order
for $N=2$ SUSY  to be preserved on the squashed four-sphere.  We have set  $v_{12}=0$, this choice of
background preserves $SU(2)\times U(1)\times U(1)_R$ symmetry. Should
we take $v_{12}\ne0$ it can be shown that the symmetry is reduced to $SU(2)\times U(1)^{\prime}$ where $U(1)^{\prime}\equiv (U(1)\times U(1)_R)_{diagonal}$.
%Regularity of the background fields
\subsection{Regularity of the Background Fields}\label{regularity}
Our metric should look like the round $S^4$  at the North and South poles, this
implies that $f(r)=h(r)=0$ at $r=0$ and $r=\pi$.
Moreover for our metric to be non-singular in the interval $\pi>r>0$, the functions $f(r)$ and $h(r)$ are strictly non-zero and do not change sign inside the interval.\\
North pole $(r=0)$: Near the North pole the regularity of invariant quantities $R$, $R_{\mu\nu}R^{\mu\nu}$ and of the background fields both in flat tangent space indices and curved space indices,  fixes $f(r)$, $h(r)$ and $s(r)$
in the following form:
\begin{align}
\begin{split}
&h(r)=r+\text{h}_{n_3} r^3 +O(r^4),\\
&f(r)=r+\text{f}_{n_3} r^3 +O(r^4),\\
&s(r)=s_{n_0}+s_{n_2} r^2+s_{n_3} r^3+O(r^4).
\end{split}
\end{align}
There are higher order terms, but those are irrelevant to the present analysis.\\
South pole$(r=\pi)$: Similarly near the South pole the regularity requirements fix $f(r)$, $h(r)$ and $s(r)$ in the following way
\begin{align}
\begin{split}
&h(r)=\pi-r+\text{h}_{s_3} (\pi -r)^3+O\left((\pi-r)^4\right), \\
&f(r)=\pi-r+\text{f}_{s_3} (\pi -r)^3 +O\left((\pi-r)^4\right),\\
&s(r)=(\pi -r) s_{s_1}+(\pi -r)^3 s_{s_3}+O\left((\pi-r)^4\right).
\end{split}
\end{align}
Where $h_{n_3},f_{n_3},s_{n_0},s_{n_2},s_{n_3},h_{s_3},f_{s_3},s_{s_1},s_{s_3}$ are arbitrary real constants.\\
 For reasons that will become clear later,
a quantity of interest which we want to calculate  is $(s(r)^2-\frac{c^2 h(r)^2}{s(r)^2})$. At the North pole it evaluates to $s_{n_0}^2$,
whereas at the South pole it evaluates to $-\frac{c^2}{s_{s_1}^2}$. So it has the interesting property that it changes sign
between North and South poles and hence passes through zero. This
result will have important consequences later on, in
section $\ref{One-loop determinant}$ when we will calculate
the one-loop determinant, where we show that the relevant differential
operators are transversally elliptic.
Before proceeding, we want to comment that there is an ambiguity in the choice of the functions $f(r)$, $h(r)$ and $s(r)$ at the North and South poles, that is, if we take following choice for these functions at the North pole
\begin{align}
\begin{split}
&h(r)=-r+\text{h}_{n_3} r^3+O(r^4), \\
&f(r)=r+\text{f}_{n_3} r^3+O(r^4) ,\\
&s(r)= s_{n_1}+s_{n_3} r^3+O(r^4),
\end{split}
\end{align}
and the following choice at the South pole
\begin{align}
\begin{split}
&h(r)=r-\pi+\text{h}_{s_3} (\pi-r)^3+O\left((\pi-r)^4\right), \\
&f(r)=\pi-r+\text{f}_{s_3} (\pi-r)^3+O\left((\pi-r)^4\right), \\
&s(r)=s_{s_0}+s_{s_2} (\pi-r)^2+s_{s_3} \left(\pi-r)^3+O((\pi-r)^4\right),
\end{split}
\end{align}
all the background fields are still regular there. The only difference is that the quantity $(s(r)^2-\frac{c^2 h(r)^2}{s(r)^2})$ evaluates to $-\frac{c^2}{s_{n_1}^2}$ at
the South pole and to $s_{s_0}^2$ at the South pole. Every other result remains the same.
\section{Multiplets}\label{multiplets}
\subsection{Vector Multiplet}
 In 4D N=2 SUSY with Eucildean signature, vector multiplets  are made
 of a gauge field $A_m$, two independent gauginos $\lambda_{\alpha
   A}$, $\bar{\lambda}_{\dot{\alpha}A}$, two scalar fields $\phi$,
 $\bar{\phi}$ and an auxiliary field $D_{AB}=D_{BA}$, all  Lie algebra
 valued.  The supersymmetric Yang-Mills Lagrangian with the additional
 couplings to the backgrounds was written in \cite{Hama:2012bg}, we write it again for completeness:
 \begin{align}
 \begin{split}
& L_{YM}=\text{Tr}[\frac{1}{2}F^{mn}F_{mn}+16F_{mn}(\bar{\phi}T^{mn}+\phi\bar{T}^{mn})+64\bar{\phi}^2T_{mn}T^{mn}+64\phi^2\bar{T}^{mn}\bar{T}_{mn} \\ & -4D_m\bar{\phi}D^m\phi+2M\phi\bar{\phi}-2\iota\lambda^A\sigma^mD_m\bar{\lambda}_A-2\lambda^A[\bar{\phi,\lambda_A}]+2\bar{\lambda}^A[\phi,\bar{\lambda}]+4[\phi,\bar{\phi}]^2-\frac{1}{2}D^{AB}D_{AB}].
 \end{split}
 \end{align}
 With the inclusion of the $\theta$-term:
 \begin{align}
 S_{YM}=\frac{1}{g^2_{YM}}\int d^4x\sqrt{g}L_{YM}+\iota\frac{\theta}{8\pi^2}\int \text{Tr}(F\wedge F).
 \end{align}
 \subsection{Hypermultiplet}
 The  hypermultiplet consists of scalars $q_{AI}$ and fermions $\psi_{\alpha A}$,
 $\bar{\psi}^{\dot{\alpha}}_I$ satisfying reality conditions  \cite{Hama:2012bg}. The index $I$ runs from
 $1$ to  $2r$. There is also
 an auxiliary scalar $F_{IA}$ transforming as a doublet under a  local $SU(2)_{\check{R}}$
 symmetry. This symmetry and the auxiliary field are introduced in the theory by the requirement that the SUSY
 algebra of matter multiplet is closed off shell  respect to the supercharge that is used to localize \cite{Pestun:2007rz}.
 From \cite{Hama:2012bg} the gauge covariant kinetic Lagrangian for the hypermultiplet is
 \begin{align}
 \begin{split}
  L_{mat}=\frac{1}{2}D_mq^AD^mq_A-q^A\{\phi,\bar{\phi}\}q_A+\frac{\iota}{2}q^AD_{AB}q^B+\frac{1} {8}(R+M)q^Aq_A-\frac{\iota}{2}\bar{\psi}\bar{\sigma}^mD_m\psi- \\ \frac{1}{2}\psi\phi\psi+  \frac{1}{2}\bar{\psi}\bar{\phi}\bar{\psi}+\frac{\iota}{2}\psi\sigma^{kl}T_{kl}\psi-\frac{\iota}{2}\bar{\psi}\bar{\sigma}^{kl}\bar{T}_{kl}\bar{\psi}-q^A\lambda_A\psi+\bar{\psi}\bar{\lambda}q^A-\frac{1}{2}F^AF_A.
 \end{split}
 \end{align}
\subsection{Closure of the Supercharge Algebra}
For localization computation we need to identify a  continuous fermionic symmetry $\textbf{Q}$ and
the corresponding Killing spinor is taken to be commuting.
The supersymmetry transformation $Q$ acting on the fields of $N=2$ SUSY theory squares into a combination of bosonic symmetries:
\begin{align}
\textbf{Q}^2\equiv  L_v+Gauge(\hat{\Phi})+Lorentz(L_{ab})+Scale({\omega})+R_{U(1)}(\Theta)+R_{SU(2)}(\hat{\Theta}_{AB})+\check{R}_{SU(2)}(\hat{\check{\Theta}}),
\end{align}
with various parameters defined as in \cite{Hama:2012bg}.
For the vector multiplet the SUSY algebra is closed off shell, the only requirement being that the Killing spinor equations be satisfied. For the hypermultiplet the closure of full $N=2$ SUSY algebra requires the existence of infinite number of auxiliary spinors and  auxiliary fields. But for localization computation we need only one supercharge corresponding to a particular Killing spinor and in this case only finite number of auxiliary spinors are required. These auxiliary spinors are required to satisfy certain constraint equations (see \cite{Pestun:2007rz}).\\
Next we  compute these transformation parameters for our background. First of all, we observe that $\xi^A\xi_{pA}=\bar{\xi}^A\bar{\xi}_{pA}=0$. This condition implies that $\omega=\Theta=0$. In other words the square of the supersymmetry transformation does not give rise to dilation or $U(1)_R$ transformation. This condition is necessary because the non-zero values of the background fields  $T_{ab}$ and $\bar{T}_{ab}$ break the $U(1)_R$ symmetry anyway.\\
The explicit expression for other transformation parameters are given below
\begin{align}
\begin{split}
L_{ab} &=\left(
\begin{array}{cccc}
 0 & -8 c & 0 & 0 \\
 8 c & 0 & 0 & 0 \\
 0 & 0 & 0 & 0 \\
 0 & 0 & 0 & 0 \\
\end{array}
\right),\\
\Theta_{AB} &=\left(
\begin{array}{cc}
 0 & 2 c \left(\frac{h(r)^2}{f(r)^2}-\frac{2 s'(r) h(r)}{s(r)}+h'(r)\right) \\
 2 c \left(\frac{h(r)^2}{f(r)^2}-\frac{2 s'(r) h(r)}{s(r)}+h'(r)\right) & 0 \\
\end{array}
\right),\\
\hat{\Theta}^A_{B} &=\left(
\begin{array}{cc}
 4 c & 0 \\
 0 & -4 c \\
\end{array}
\right),\\
Lie_v\xi &=\left(
\begin{array}{cc}
 -\frac{2 c s(r) \left(\left(h'(r)-2\right) f(r)^2+h(r)^2\right)}{f(r)^2} & 0 \\
 0 & \frac{2 c s(r) \left(\left(h'(r)-2\right) f(r)^2+h(r)^2\right)}{f(r)^2} \\
 \frac{2 \iota c^2 h(r) \left(f(r)^2 \left(h'(r)+2\right)-h(r)^2\right)}{f(r)^2 s(r)} & 0 \\
 0 & \frac{2 \iota c^2 h(r) \left(f(r)^2 \left(h'(r)+2\right)-h(r)^2\right)}{f(r)^2 s(r)} \\
\end{array}
\right),
\end{split}
\end{align}
where the Lie derivative $Lie_v$ is defined as $L_v\xi\equiv \upsilon^m D_m\xi+\frac{1}{4}D_{[a}\upsilon_{b]}\Gamma^{ab}\xi$. The non-zero $L_{ab}$ implies the fact
that the $U(1)$ group which is used to find the fixed points of the manifold, belongs to the
Cartan of $SU(2)$ part of the isometry group $SU(2)\times U(1)$. Therefore it follows that our
Killing spinor is invariant under $Q^2$. In 4-component notation:
\begin{align}
Q^2\xi= \iota Lie_v\xi-\xi\hat{\Theta}=0.
\end{align}
%The background fields $V_{mAB},T_{kl},\bar{T}_{kl},M$ are also invariant under
%$Q^2$ since they are constructed from $Lie_v$-invariant functions and Killing spinor.\\
%To determine the action of $Q^2$ on all the fields, we still need to determine
%$\check{\xi}$ and the background $SU(2)_{\check{R}}$ gauge field
%$\check{V}_{mAB}$.
The auxiliary spinor, which helps to close off-shell the supersymmetry,
is given by:
\begin{align}
\begin{split}
%&\check{\xi}_A =-\frac{c h(r)}{s(r)^2}\xi_A\\
%&\bar{\check{\xi}}_A = \frac{s(r)^2}{c h(r)}\bar{\xi}_A\\
&\check{\xi} =\left(
\begin{array}{cc}
 \frac{c h(r)}{s(r)} & 0 \\
 0 & \frac{c h(r)}{s(r)} \\
 -\iota s(r) & 0 \\
 0 & \iota s(r) \\
\end{array}
\right).
\end{split}
\end{align}
 %We will take the following choice for the   $SU(2)_{\check{R}}$ guage field $\check{V}_{mAB}$ .
%\begin{align}
%\check{V}_{mAB}=V_{mAB}
%\end{align}
To fix the background $SU(2)_{\check{R}}$, we have to fix the corresponding gauge field $\check{V}_m$:
\begin{align}
\begin{split}
&\check{V}_m =\left(
\begin{array}{cc}
 \iota \check{v}_{3m} & \iota (\check{v}_{1m}+\iota \check{v}_{2m}) \\
 \iota (\check{v}_{1m}-\iota \check{v}_{2m}) & -\iota \check{v}_{3m} \\
\end{array}
\right).
\end{split}
\end{align}
The requirement that all the background fields be invariant under the action of
$\textbf{Q}^2$ fixes all the  components of $\check{V}_m$ to zero except $\check{v}_{33},\check{v}_{34}$,
which remain arbitrary.\\
After the gauge fixing, $\hat{\check{\Theta}}^A_{B}$ becomes
\begin{align}
\hat{\check{\Theta}}^A_{B}=\left(
\begin{array}{cc}
 -4 (h(r) \check{v}_{33}(r) c+c) & 0 \\
 0 & 4 (h(r) \check{v}_{33}(r) c+c) \\
\end{array}
\right).
\end{align}
And also  the auxiliary spinor $\check{\xi}$ is proven to be invariant under $Q^2$
\begin{align}
\textbf{Q}^2\check{\xi}=0.
\end{align}
\section{Localization}\label{path integration}
\subsection{$S_{YM}$ Saddle Points  }
 The path integral computation of the expectation value of an
observable of a supersymmetric $YM$ theory which is invariant under a supercharge $\textbf{Q}$ localizes to a subset $S_{Q}$ of
the entire field space. The zero locus of the supercharge $\textbf{Q}$ coincides with the set of bosonic configurations
for which the supersymmetry variations of the fermions vanish:  
\begin{align}
\textbf{Q}\Psi=0\quad \text{for all fermions}\quad \Psi.
\end{align}
This is easily seen if  we can take as regulator the \textbf{Q}-exact deformation:
$\textbf{Q}V=\textbf{Q}((\textbf{Q}\Psi)^{\dag}\Psi)$. \\
To take into account the gauge fixing, the superchage $\textbf{Q}$ is generalized to $\hat{\textbf{Q}}\equiv\textbf{Q}+\textbf{Q}_B$, where
$\textbf{Q}_B$ is the BRST-supercharge. However as pointed out in  \cite{Pestun:2007rz}, this does not affect the zero locus.
To effectively  calculate the zero locus of the supercharge,  we  add to the Lagrangian a $\textbf{Q}$-exact quantity
$\textbf{Q}V$, whose critical point set is  $S_{Q}$ and whose bosonic part is semi-positive definite.
Now either solving the localization equation
\begin{align}
\hat{\textbf{Q}}\lambda=0,
\end{align}
directly or analyzing
  the $\hat{\textbf{Q}}$-transform of the following quantity,
\begin{align}
\label{eq:QV}
 V=\text{Tr}[(\hat{\textbf{Q}}\lambda_{\alpha A})^{\dagger}\lambda_{\alpha A}+(\hat{\textbf{Q}}\bar{\lambda}^{\dot{\alpha}}_A)^{\dagger}\bar{\lambda}^{\dot{\alpha}}_A],
\end{align}
which has  semi-positive definite bosonic part. In writing explicitly
\eqref{eq:QV} we use the proper reality conditions which make the
action well defined. We get the analogous expression to the equation
(4.2) in \cite{Hama:2012bg}. \\
Analyzing that expression we get the following   partial differential equations for
$\phi-\bar{\phi}\equiv \phi_2(\psi,\theta,\varphi,r)$, where we make
use of Bianchi identities to get the second equation: 
\begin{align}
\partial_{\psi}\phi_2(\psi,\theta,\varphi,r)=0,
\end{align}
and
\begin{eqnarray}
\tilde{\nabla}^2\phi_2(\theta,\varphi,r)+
\frac{f(r)^2}{2h(r)}\xi\Gamma^m\xi_p \partial_m\phi_2(\theta,\varphi,r)+G(r)\phi_2(\theta,\varphi,r)=0,
\end{eqnarray}
where in the second equation we used the fact that $\phi_2(\psi,\theta,\varphi,r)$ is independent of
$\psi$-coordinate.
 $\tilde{\nabla}^2$ is the following  Laplacian like operator:
\begin{eqnarray}
\tilde{\nabla}^2 *=\frac{f(r)^2}{2 h(r)}\frac{h(r)}{\sqrt{g}f(r) \xi_n}\nabla_\mu\Bigl(\sqrt{g}
\xi^2_ng^{\mu \nu}\nabla_\nu(\frac{f(r)}{h(r)}*)\Bigr)
\end{eqnarray} 
$\xi_n=\xi.\xi$ is the proper norm of the four component spinor and $G(r)$
\begin{eqnarray}
G(r)&=&\frac{1}{h(r)^3 s(r)^3}\Biggl(-c^2 h(r)^4 \left(s(r)
  \left(f'(r)^2+2 h'(r)\right)-2 f(r) f'(r) s'(r)\right)\Biggr.\nonumber\\
 &-&h(r)^2
\left(-3 c^2 f(r)^2 s(r) h'(r)^2+2 f(r) s(r)^4 f'(r) s'(r)+s(r)^5
  f'(r)^2\right) \nonumber\\  &+&h(r)^3 \left(c^2 f(r)^2 s(r) h''(r)+2
  s'(r) \left(s(r)^4-2 c^2 f(r)^2 h'(r)\right)\right)+2 c^2 h(r)^5
s'(r)\nonumber\\
&+&f(r) h(r) s(r)^4 \left(2 h'(r) \left(s(r) f'(r)+2 f(r)
    s'(r)\right)+f(r) s(r) h''(r)\right) \nonumber\\ \Biggl.&-&f(r)^2 s(r)^5 h'(r)^2\Biggr)
\end{eqnarray}
For the round sphere
\begin{align}
\begin{split}
f(r)=\sin{r},\quad h(r)=\sin{r},\quad
s(r)=\frac{1}{\sqrt{2}}\cos(\frac{r}{2}),
\end{split}
\end{align}
  the field $\phi_2=0$ at the localization locus, which  will also ensure that $A_m=0$ at the locus. 
This result  is  true in an open neighborhood of the round $S^4$, as
appears also in \cite{Hama:2012bg}, and so we will assume it is the
solution to the locus equations. \\
The saddle points are thus labeled by a Lie Algebra valued constant $a_0$, and are given by the equations\cite{Pestun:2007rz,Hama:2012bg}:
\begin{align}
 A_m=0, \qquad \phi=\bar{\phi}= a_0,\qquad D_{AB}=-\iota a_0 \omega_{AB},
\end{align}
The value of the Super-Yang-Mills action on this saddle point is then:
\begin{align}
\frac{1}{g^2_{YM}}\int d^4x\sqrt{g}L_{YM}|_{saddle point}=\frac{2 \pi ^3 \text{Tr}\left[\text{a}_0^2\right]}{c^2\text{g}_{YM}^2}.
\end{align}
\subsection{Saddle points for Matter multiplet}
To find the saddle points of the matter multiplet we will use the following fermionic functional
\begin{align}
 V_{mat}=\text{Tr}[(\hat{\textbf{Q}}\psi_{\alpha I})^{\dagger}\psi_{\alpha I}+(\hat{\textbf{Q}}\bar{\psi}^{\dot{\alpha}}_I)^{\dagger}\bar{\psi}^{\dot{\alpha}}_I].
\end{align}
The bosonic part of $\hat{\textbf{Q}}V_{mat}$ is
\begin{align}
\hat{\textbf{Q}}V_{mat}|_{bos}=\text{Tr}[(\hat{\textbf{Q}}\psi_{\alpha I})^{\dagger}\hat{\textbf{Q}}\psi_{\alpha I}+(\hat{\textbf{Q}}\bar{\psi}^{\dot{\alpha}}_I)^{\dagger}\hat{\textbf{Q}}\bar{\psi}^{\dot{\alpha}}_I].
\end{align}
%Following \cite{Pestun:2007rz} 
It is easy to check that:
\begin{align}
\hat{\textbf{Q}}V_{mat}|_{bos}=4
\| \xi\|^2(\frac{1}{2}(D_{m}q^{AI}-P_{m}q^{AI})^2+M_q(r)q^{AI}q_{IA}-\frac{1}{2}F^{AI}F_{IA}),
\end{align}
where
\begin{align}
P_{m A}^{ B}=\frac{1}{\| \xi\|^2}(2 (\epsilon\xi\gamma_m\xi_p+\epsilon\xi T\gamma_m\xi)^B_{\quad A}+D^n
\text{Log}(\| \xi\|^2)(\epsilon\xi\gamma_{nm}\xi)^B_{\quad A}),
\end{align}
and
\begin{align}\begin{split}
M_q&=-\frac{1}{4}R+\frac{1}{\| \xi\|^2}(8\xi^A_p\xi_{pA}+\xi^A\gamma^mT^2\gamma_m\xi_A-D^n
\text{Log}(\| \xi\|^2)\xi^A(3\gamma_m\xi_{pA}+T\gamma_m\xi_A)+\\ & \frac{1}{2}( P^{mA}_{\qquad B}P_{mA}^B)) -\frac{1}{2\| \xi\|^2}P^{mA}_{\quad A}P^{B}_{m B},
\end{split}
\end{align}
where $\xi_A=(\xi_{\alpha A},\bar{\xi}_{\dot{\alpha}A})$,$\epsilon^{AB}$ is the $SU(2)_R$ tensor and $R$ is the Ricci scalar.
 As a result of the condition $F_{IA}^{\dagger}=-F^{AI}$ which is imposed along the contour of  path integration, all the bosonic terms are manifestly positive definite, except the term containing $M_q(r)$. For the round $S^4$
\begin{align}
M_q(r)=\frac{7}{8}+\frac{\cos (2r)}{8},
\end{align}
and it is bounded from below by $\frac{3}{4}$. Therefore there is a large open neighborhood of  the
round sphere for which $M_q(r)$ is positive definite.
So we get the  result for the saddle points of the hypermultiplet as
 \begin{align}
 q_{IA}=0,\qquad F_{IA}=0.
 \end{align}
Hence there will be no  classical contribution from the hypermultiplet  sector.
\section{One-loop determinant}\label{One-loop determinant}
To calculate the one-loop determinant we have to first fix the gauge. 
We choose the following gauge function\cite{Hama:2012bg}.
\begin{align}
 G=\iota \partial_mA^m+\iota L_v((\xi^A\xi_A-\bar{\xi}_A\bar{\xi}^A)\phi_2-\upsilon^mA_m).
\end{align}
The saddle point conditions do not change under the new supercharge $\hat{Q}^2\equiv (Q_B+Q)^2$, with the zero mode of   $\phi_1=a_0$ at the saddle point.
\subsection{Vector multiplet contribution}
The basic idea of localization is that the actual value of the path integral  or any other $Q$-closed observable remains unchanged under any $\hat{\textbf{Q}}$-exact deformation $L\rightarrow L+s\hat{\textbf{Q}}(V+V_{GF})$. By choosing the bosonic part of  $L\rightarrow L+s\hat{\textbf{Q}}(V+V_{GF})$ positive definite and sending $s\rightarrow\infty$, Gaussian approximation becomes exact for the path integral over the fluctuations around the locus. The Gaussian integral evaluates to the square root of the ratio between the determinant of a fermionic kinetic operator $K_{fermion}$ and that of a bosonic kinetic operator $K_{boson}$. These kinetic operators coming from the quadratic part of the
 $\hat{\textbf{Q}}$-exact regulator.\\
To compute the 1-loop contribution it  is convenient to change variables in the
path integral to a set,  $X$,$\Xi$, which makes manifest the cohomology of $\hat{\mathbf{Q}}$ \cite{Pestun:2007rz,Hama:2012bg} . After doing that,  the quadratic part of $V+V_{GF}$ can be written as:
\begin{align}\label{eq:Vquad}
 (V+V_{GF})|_{quadratic}=(\hat{\textbf{Q}}\textbf{X},\Xi)\left(
\begin{array}{cc}
 \text{D}_{00} & \text{D}_{01} \\
 \text{D}_{10} & \text{D}_{11} \\
\end{array}
\right)\left(
\begin{array}{c}
 \textbf{X} \\
 \hat{\textbf{Q}}\Xi \\
\end{array}
\right),
\end{align}
where $D_{ij}$ are differential operators and $X$,$\Xi$ are cohomologically paired bosonic and fermonic fields respectively,
\begin{align}
\Xi \equiv (\Xi_{AB},\bar{C},C),\qquad \textbf{X}=(\phi_2,A_m;\bar{a}_0,B_0),
\end{align}
and
\begin{align}
\Xi_{AB}\equiv 2 \bar{\xi}_{(A}\bar{\lambda}_{B)}-2 \xi_{(A}\lambda_{B)},
\end{align}
where $\bar{C},C,\bar{a}_0,B_0$ belong to the ghost multiplets 
%as introduced in \cite{Pestun:2007rz,Hama:2012bg} 
%for fixing the local gauge redundancy of the path integral. 
The fields $X$ and $\Xi$ can be regarded as sections of bundles $E_0$, $E_1$ over the squashed sphere and hence $D_{10}$ acts on the complex as  $D_{10}: \Gamma(E_0)\rightarrow \Gamma(E_1)$. The invariance of the deformation term $\hat{Q}(V+V_{GF})$ under the action
of $\hat{Q}$ and the pairing of the fields under  $\hat{\textbf{Q}}^2=\textbf{H}$ leads to the cancellations between bosonic and fermionic
fluctuations, which gives the  following ratio \cite{Pestun:2007rz,Hama:2012bg}:
\begin{align}\label{eq:1-loop}\frac{\textbf{det}_{CokerD_{10}}\textbf{H}}{\textbf{det}_{KerD_{10}}\textbf{H}}.
\end{align}
The fact that $\hat{\textbf{Q}}^2$ commutes with the differential operators $D_{ij}$ is used in the derivation of the last expression and is a result of the invariance of $(V+V_{GF})$ under $\hat{Q}^2$. This can readily be seen by considering $\hat{Q}^2(V+V_{GF})_{Quad}$.
\begin{align}
\hat{Q}(V+V_{GF})_{Quad}=\left(
\begin{array}{cc}
 X &   \hat{Q}\Sigma \\
\end{array}
\right)\mathbb{D}\left(
\begin{array}{cc}
 -\hat{Q}^2 & 0 \\
 0 & 1 \\
\end{array}
\right)\left(
\begin{array}{c}
 X \\
   \hat{Q}\Sigma \\
\end{array}
\right)-\left(
\begin{array}{cc}
  \hat{Q}X & \Sigma  \\
\end{array}
\right)\mathbb{D}\left(
\begin{array}{cc}
 1 & 0 \\
 0 & \hat{Q}^2 \\
\end{array}
\right)\left(
\begin{array}{c}
  \hat{Q}X \\
 \Sigma  \\
\end{array}
\right),
\end{align}
where $\mathbb{D}\equiv \left(
\begin{array}{cc}
 \text{D}_{00} & \text{D}_{01} \\
 \text{D}_{10} & \text{D}_{11} \\
\end{array}
\right)$.
\\Then
\begin{align}\begin{split}
\hat{Q}^2(V+V_{GF})_{Quad}&=\left(
\begin{array}{cc}
 \hat{Q} X &   \hat{Q}^2\Sigma \\
\end{array}
\right)\left(
\begin{array}{cc}
 -\hat{Q}^2 & 0 \\
 0 & 1 \\
\end{array}
\right)\mathbb{D}\left(
\begin{array}{c}
 X \\
   \hat{Q}\Sigma \\
\end{array}
\right)+\left(
\begin{array}{cc}
 X &   \hat{Q}\Sigma \\
\end{array}
\right)\left(
\begin{array}{cc}
 -\hat{Q}^2 & 0 \\
 0 & 1 \\
\end{array}
\right)\mathbb{D}\left(
\begin{array}{c}
 \hat{Q} X \\
   \hat{Q}^2\Sigma \\
\end{array}
\right)-\\& \left(
\begin{array}{cc}
 \hat{Q}^2 X &   \hat{Q}\Sigma \\
\end{array}
\right)\mathbb{D}\left(
\begin{array}{cc}
 1 & 0 \\
 0 & -\hat{Q}^2 \\
\end{array}
\right)\left(
\begin{array}{c}
  \hat{Q}X \\
 \Sigma  \\
\end{array}
\right)+\left(
\begin{array}{cc}
 \hat{Q} X & \Sigma  \\
\end{array}
\right)\mathbb{D}\left(
\begin{array}{cc}
 1 & 0 \\
 0 & \hat{Q}^2 \\
\end{array}
\right)\left(
\begin{array}{c}
  \hat{Q}^2X \\
   \hat{Q}\Sigma \\
\end{array}
\right)\\ &= \left(
\begin{array}{cc}
  \hat{Q}X & \Sigma  \\
\end{array}
\right)\left(
\begin{array}{cc}
 1 & 0 \\
 0 & -\hat{Q}^2 \\
\end{array}
\right)\left(
\begin{array}{cc}
 -\hat{Q}^2 & 0 \\
 0 & 1 \\
\end{array}
\right)\mathbb{D}\left(
\begin{array}{c}
 X \\
   \hat{Q}\Sigma \\
\end{array}
\right)+\left(
\begin{array}{cc}
 X &   \hat{Q}\Sigma \\
\end{array}
\right)\left(
\begin{array}{cc}
 -\hat{Q}^2 & 0 \\
 0 & 1 \\
\end{array}
\right)\\& \mathbb{D}\left(
\begin{array}{cc}
 1 & 0 \\
 0 & \hat{Q}^2 \\
\end{array}
\right)\left(
\begin{array}{c}
  \hat{Q}X \\
 \Sigma  \\
\end{array}
\right)-\left(
\begin{array}{cc}
 X &   \hat{Q}\Sigma \\
\end{array}
\right)\left(
\begin{array}{cc}
 -\hat{Q}^2 & 0 \\
 0 & 1 \\
\end{array}
\right)\mathbb{D}\left(
\begin{array}{cc}
 1 & 0 \\
 0 & \hat{Q}^2 \\
\end{array}
\right)\left(
\begin{array}{c}
  \hat{Q}X \\
 \Sigma  \\
\end{array}
\right)+\\ &\left(
\begin{array}{cc}
 \hat{Q} X & \Sigma  \\
\end{array}
\right)\mathbb{D}\left(
\begin{array}{cc}
 1 & 0 \\
 0 & \hat{Q}^2 \\
\end{array}
\right)\left(
\begin{array}{cc}
 \hat{Q}^2 & 0 \\
 0 & 1 \\
\end{array}
\right)\left(
\begin{array}{c}
 X \\
   \hat{Q}\Sigma \\
\end{array}
\right).
\end{split}
\end{align}
Now with the requirement that $[\hat{Q}^2,D_{ij}]=0$, different terms cancel among each other and we get
\begin{align}
\hat{Q}^2(V+V_{GF})_{Quad}=0.
\end{align}
\subsection{Index of $D_{10}$}
To evaluate the ratio \eqref{eq:1-loop} through the index computation, we first note that  the constant fields $B_0$, $\bar{a}_0$ have each weight  $0$ under the action of  $U(1)$ at the poles and  are thus regarded as sitting
in the kernel of $D_{10}$ and  making a contribution of $2$. For the contribution of other fields we need an explicit expression for $D_{10}$
\footnote{Strictly speaking the relevant differential operator for the index computation  is a combination of the original $D_{10}$ and $D_{11}$. But it turns out that this operator commutes with $\textbf{H}$ and  the distinction becomes irrelevant. }, which is read from equation \eqref{eq:Vquad}
%\begin{align}
%\Xi D_{10}X+\Xi D_{11}\hat{\textbf{Q}}\Xi=\text{Tr}[\bar{c}G-D_mc(\hat{\textbf{Q}}\Psi^m)^{\dagger}+\frac{1}{2}\Xi_{AB}(\hat{\textbf{Q}}\Xi_{AB})^{\dagger}]|_{quad},
%\end{align}
%given in \cite{Hama:2012bg}, where
%\begin{align}
%\Psi_m\equiv \textbf{Q}A_m.
%\end{align}
%where ignoring any non-linear term:
%\begin{align}
%\begin{split}
%&(\hat{\textbf{Q}}\Psi_m)^{\dagger}=-\iota L_vA_m+D_m(\hat{\Phi}-2\iota(\xi^A\xi_A-\bar{\xi}_A\bar{\xi}^A)\phi_2
%2\iota \upsilon^nA_n),\\
%&(\hat{\textbf{Q}}\Xi_{AB})^{\dagger}=-\xi^A\sigma^{kl}\xi^B(F_{kl}-8\phi T_{kl}+
%8\bar{\phi S_{kl}})+\bar{\xi}^A\bar{\sigma}^{kl}\bar{\xi}^B(F_{kl}-8\bar{\phi}\bar{T}_{kl}+8\phi \bar{S}_{kl})-4\xi^{(A}\sigma^n\bar{\xi}^{B)}D_n\phi_2-D^{AB}
%\end{split}
%\end{align}
%$\hat{\textbf{Q}}\Psi^m$ and $\hat{\textbf{Q}}\Xi_{AB}$ can easily be calculated by using the SUSY transformation rules of vectormultiplet fields.
To compute the index of $D_{10}$ it is better to use its, symbol $\sigma(D_{10})$, this  is computed by taking  the Fourier transform of the operator $D_{10}$  and then retaining only the highest order derivative (momentum) terms \cite{Pestun:2007rz}.
To write the symbol explicitly we have to express the Fourier transform of $D_{10}$ in the following orthonormal basis of  four unit vector fields $\mu^m_a~(a=1,2,3,4)$ , which relabels the original vielbein basis
\begin{align}
\begin{split}
-2\iota (\tau^a)^A_B\bar{\xi}^B\bar{\sigma}^m\xi_A=4 c h(r)\mu^m_a,\qquad 2\bar{\xi}^A\bar{\sigma}^m\xi_A=4 c h(r)\mu^m_4,\quad(a=1,2,3),
\end{split}
\end{align}
 Here $c$ is the constant appearing in the definition of the Killing spinor. 
 %Using the proper generalization of $\hat{\textbf{Q}}\Psi^m$ and $\hat{\textbf{Q}}\Xi_{AB}$ \cite{Hama:2012bg}   to our squashed $S^4$ background in  this basis we find for the symbol
So the symbol  is given by: 
\begin{eqnarray}
 \sigma(D_{10})=\left(
\begin{array}{ccccc}
 p_4 W(r) &p_3 & -p_2 & -p_1 W(r) & -4 c p_1h(r) \\
 -p_3& p_4 W(r) & p_1 & -p_2 W(r) & -4 c p_2 h(r) \\
 p_2 & -p_1 & p_4 W(r) & -p_3 W(r) & -4 c p_3 h(r) \\
p p_1 p_4 & p_2 p_4 & p_3 p_4 & p_4^2-8 c \left(p_1^2+p_2^2+p_3^2\right) h(r) & 2
   \left(p_1^2+p_2^2+p_3^2\right) W(r) \\
\end{array}
\right),
\end{eqnarray}
where $W(r)\equiv 2 s(r)^2-\frac{2 c^2 h(r)^2}{s(r)^2}$. 
This matrix can be block diagonalized  in terms of 1$\times$1 and $4\times4$ factors, the relevant part of the symbol  to compute  the index is  the following $4\times4 $ block ,
\begin{align}
\sigma(D^{\prime}_{10})=\left(
\begin{array}{cccc}
 p_4 W(r) &p_3 & -p_2 & -p_1 \\
 -p_3 & p_4W(r) & p_1 & -p_2 \\
 p_2 & -p_1 & p_4 W(r) & -p_3 \\
 p_1 & p_2 & p_3& p_4 W(r) \\
\end{array}
\right).
\end{align}
%An elliptic differential operator is one  whose symbol is invertible for non-zero four vector $(p_1,p_2,p_3,p_4)$.
 The determinant of this symbol is:
\begin{align}
\text{Det}(\sigma(D^{\prime}_{10}))=\left(\frac{4 c^4 p_4^2 h(r)^4}{s(r)^4}-8 c^2 p_4^2 h(r)^2+p_1^2+p_2^2+p_3^2+4 p_4^2 s(r)^4\right)^2.
\end{align}
For $p_1=p_2=p_3=0$ and $p_4\not=0$, this value of determinant changes sign between North and South poles  as discussed in section \ref{regularity}, hence it has at least one zero.  Therefore the symbol is not invertible at the location of that
zero and by definition $D_{10}$ cannot be elliptic. But  restricting the momentum to  $p_4=0$,
$\sigma$ is always invertible provided $(p_1,p_2,p_3)$ are not all zero simultaneously. Therefore  $D_{10}$ is a transversally elliptic operator with  respect to the symmetry generated by $\upsilon$.  In general the kernel and cokernel of such transversally elliptic operator are infinite dimensional, but since $[\hat{Q}^2,D_{ij}]=0$,  they can both be splitted into irreps. of $\textbf{H}$ with finite multiplicities, these multiplicities can be read off from the index theorem as explained in \cite{Pestun:2007rz}.
The index theorem localizes the contributions to the fixed points of the action of $\textbf{H}$, that is to the North and South poles of the squashed $S^4$. According to the Atiyah-Bott \cite{1967} formula, the index is given by,
\begin{align}
ind(D^{\prime}_{10})=\sum_{\text{x=fixed points}}\frac{\text{Tr}_{E_0}(\gamma)-\text{Tr}_{E_1}(\gamma)}{\text{det}(1-\frac{\partial\tilde{x}}{\partial x})},
\end{align}
where $\gamma$ denotes the eigenvalue of the action of the operator
$e^{\iota \textbf{H} t}$ on the vector and $SU(2)_R$ indices of the
fields. So we need the action of $e^{\iota \textbf{H} t}$  Near the North  and
South poles, on the local coordinates $z_1\equiv x_1+\iota x_2$,$z_2\equiv x_3+\iota x_4$, where  we are defining near the North pole:
\begin{align}
\begin{split}
x_1+\iota x_2=r \cos \left(\frac{\theta }{2}\right)e^{\iota\frac{\psi+\varphi}{2}},\\
x_3+\iota x_4=r \sin \left(\frac{\theta }{2}\right)e^{\iota\frac{\psi-\varphi}{2}},
\end{split}
\end{align}
so,
\begin{align}
z_1\to e^{4 \iota c t} z_1\equiv q_1 z_1,\qquad z_2\to e^{4\iota c t}z_2\equiv q_2 z_2,
\end{align}
 With$0\le \theta \le \pi$, $0\le \phi \le 2 \pi$, $0\le \psi \le 4
\pi$ % The determinant gives
%\begin{align}
%{\text{det}(1-\frac{\partial\tilde{x}}{\partial x})}=(1-q)^2(1-\bar{q})^2.
%\end{align}
As for  the action of $Q^2$ on  the fields of vector
multiplet,  its eigenvalues turn out to be of the same form as in
\cite{Hama:2012bg}, except that in our case $q_1=q_2=q=e^{4 \iota c
  t}$. Putting all  together, also the similar contribution from the South pole, we get the index $D_{10}$.\\
The one loop determinant can be  computed by extracting the spectrum
of eigenvalues of $\textbf{H}$ from the index. For a non-abelian group
G, with $a_0$ in its Cartan sub algebra, the one loop contribution of
the vector multiplet can be written as \cite{Hama:2012bg}:
\begin{align}
\begin{split}
Z^{vec}_{1-loop}=(\frac{detK_{fermion}}{detK_{boson}})^{\frac{1}{2}}
=\prod_{\alpha\in\Delta_{+}}\frac{1}{(\hat{a}_0.\alpha)^2}\prod_{m,n\ge0}((m+n)+\iota \hat{a}_0.\alpha)((m+n+2)+\iota \hat{a}_0.\alpha)\\((m+n)-\iota \hat{a}_0.\alpha)((m+n+2)-\iota \hat{a}_0.\alpha)\\=\prod_{\alpha\in\Delta_{+}}
\frac{\Upsilon_{1}(\iota \hat{a}_0.\alpha)\Upsilon_{1}(-\iota \hat{a}_0.\alpha)}{(\hat{a}_0.\alpha)^2},
\end{split}
\end{align}
where $\hat{a}_0\equiv \frac{a_0}{4c}$. The function $\Upsilon(x)$ has zeros at $x=-(m+n),(m+n+2)$, this function is implemented
to regularized the infinite products. It is defined by:
\begin{align}
\Upsilon_b(x)=\prod_{n_1,n_2\ge 0}(b n_1+\frac{n_2}{b}+x)(b n_1+\frac{n_2}{b}+b+\frac{1}{b}-x),
\end{align}
where $b$ is a constant that in the case of \cite{Hama:2012bg} is exactly the squashing parameter, while and in our case $b=1$.
 \subsection{Hypermultiplet one-loop contribution }
We begin also with cohomological pairing \cite{Pestun:2007rz,Hama:2012bg}  for the matter sector, the computation of the one-loop determinant reduces to that of the index of an operator $D^{mat}_{10}$. This operator corresponds to the terms
bilinear in the fields $\Xi$ and $q_{IA}$ in the functional $V_{mat}$. Its symbol $\sigma(D^{mat}_{10})$ is given by
\begin{align}
\sigma(D^{mat}_{10})=\left(
\begin{array}{cc}
 \frac{2 \left((p_3-\iota p_4) s(r)^4+c^2 h(r)^2 (p_3+\iota p_4)\right)}{s(r)^4+c^2 h(r)^2} & 2 (p_1+\iota p_2) \\
 2 ((p_1-\iota p_2) & -\frac{2 \left((p_3+\iota p_4) s(r)^4+c^2 h(r)^2 (p_3-\iota p_4)\right)}{s(r)^4+c^2 h(r)^2} \\
\end{array}
\right).\end{align}
The determinant of this symbol is
\begin{align}
Det[\sigma(D^{mat}_{10})]=-\frac{4 \left(s(r)^4-c^2 h(r)^2\right)^2}{\left(c^2 h(r)^2+s(r)^4\right)^2} p_4^2-4 p_1^2-4p_2^2-4p_3^2.
\end{align}
For $p_1=p_2=p_3=0 ,p_4\ne 0$, the determinant changes sign somewhere between North and South poles (see section \ref{regularity}) and hence it possesses at least one zero. Therefore  the operator $D^{mat}_{10}$ is  again transversally elliptic with respect to the isometry generated by $L_v$ in the $p_4$ direction.\\
The index for the action of $\textbf{H}$ on different fields at the poles can be calculated by using Atiyah-Bott formula. With $q_1=q_2=e^{4 \iota c t}$ in our case of squashed $S^4$,
the eigenvalues for the action of $Q^2$ on the matter multiplet case again turn out to have the same form as in \cite{Hama:2012bg}.
%These weights can easily be read from the $Q^2$ SUSY transformation parameters
%at the two poles of the squashed $S^4$, the operator $D^{mat}_{10}$ reduces to the Dirac operator
%$D_{Dirac}$ which maps the space of positive chirality spinors $S^{+}$ to negative chirality spinor $S^{-}$
%\begin{align}
%D_{Dirac}:S^{+}\rightarrow S^{-}
%\end{align}
%its index is
%\begin{align}\begin{split}
%\text{ind}(D_{Dirac},q_1,q_2)=\frac{(q_1^{\frac{1}{2}}q_2^{\frac{1}{2}}+\bar{q}_1^{\frac{1}{2}}\bar{q}_2^{\frac{1}{2}})-(q_1^{\frac{1}{2}}\bar{q}_2^{\frac{1}{2}}+\bar{q}_1^{\frac{1}{2}}q_2^{\frac{1}{2}})}{(1-q_1)(1-q_1^{-1})(1-q_2)(1-q_2^{-1})}
%=\frac{q_1^{\frac{1}{2}}q_2^{\frac{1}{2}}}{(1-q_1)(1-q_2)}
%\end{split}
%\end{align}

For the hypermultiplets coupled to gauge symmetry, in the representation $R\bigoplus\bar{R}$
%, the total index with the contributions from North and South poles, is given by
%\begin{align}\begin{split}
%\text{ind}(D^{mat}_{10},q_1,q_2)
%=-\sum_{\rho\in R}(e^{t a_0.\rho}+e^{-t a_0.\rho})(\frac{q_1^{\frac{1}{2}}q_2^{\frac{1}{2}}}{(1-q_1)(1-q_2)}+\frac{q_1^{\frac{1}{2}}q_2^{\frac{1}{2}}}{(1-q_1)(1-q_2)})\\
%=-\sum_{\rho\in R}(e^{t a_0.\rho}+e^{-t a_0.\rho})\sum_{m,n\ge 0}(q^{m+\frac{1}{2}}_1q^{n+\frac{1}{2}}_2+q^{-m-\frac{1}{2}}_1q^{-n-\frac{1}{2}}_2)
%\end{split}
%\end{align}
%where $\rho$ runs over all the weights in a given representation and the minus sign corresponds to the fact that for the hypermultiplet, the chirality of the complex is opposite to the chirality of the $U(1)$ rotation near each of the fixed points \cite{Pestun:2007rz}. \\
the final result for the  one-loop determinant for the hypermultiplets becomes: 
\begin{align}
\begin{split}
Z^{hyp}_{1-loop}=\prod_{\rho\in R}\prod_{m,n\ge 0}((m+n+1)-\iota \hat{a}_0.\alpha)^{-1}((m+n+1)+\iota \hat{a}_0.\alpha)^{-1}\\
=\prod_{\rho\in R}\Upsilon_{1}(\iota \hat{a}_0.\rho+1)^{-1}.
\end{split}
\end{align}
where $\rho$ runs over all the weights in a given representation.
\section{Instanton contribution}\label{instanton}
Near the North pole the Killing spinor  evaluates to
\begin{align}
\xi=\left(
\begin{array}{cc}
 s_{n_0} & 0 \\
 0 & s_{n_0} \\
 \frac{\iota c r}{s_{n_0}} & 0 \\
 0 & -\frac{\iota c r}{s_{n_0}} \\
\end{array}
\right),
\end{align}
so that $\xi^A\xi_A=2 s_{n_0}^2$ and $\bar{\xi}_A\bar{\xi}^A= \frac{2c^2 r^2}{ s_{n_0}^2}$. Since $\bar{\xi}_A\bar{\xi}^A\rightarrow 0$ at the North pole,  the localization equation has to be evaluated away from the North pole  to have smooth gauge field configurations.

Similarly near the South pole
\begin{align}
\xi=\left(
\begin{array}{cc}
 (\pi -r) s_{s_1} & 0 \\
 0 & (\pi -r) s_{s_1} \\
 \frac{\iota c}{s_{s_1}} & 0 \\
 0 & -\frac{\iota c}{s_{s_1}} \\
\end{array}
\right),
\end{align}
and $\xi^A\xi_A=2 (\pi-r)^2s_{s_1}^2$ and $\bar{\xi}_A\bar{\xi}^A= \frac{2c^2 }{ s_{s_1}^2}$. In this case
$\xi^A\xi_A\rightarrow 0$. Therefore South pole has also to be excluded if smooth gauge field configurations are assumed.

To include the contribution from the poles, we first notice that because $\bar{\xi}_A\bar{\xi}^A\rightarrow 0$  at the North pole, in general $F^{+}_{mn}\ne 0,F^{-}_{mn}=0$  there and still solve the localization equation. These configurations are the pointlike anti-instantons contribution. 

Also at the North pole the following condition is satisfied for
our background
\begin{align}
\frac{1}{4}\Omega_m^{ab}\sigma_{ab}\xi_A+\iota \xi_B V^{B}_{mA}=0.
\end{align}
 
 Likewise, at the South pole $\xi^A\xi_A\rightarrow 0$, and we get the point instanton contribution $F^{+}_{mn}= 0,F^{-}_{mn}\ne 0$ and the following twisting condition is satisfied
 \begin{align}
\frac{1}{4}\Omega_m^{ab}\bar{\sigma}_{ab}\bar{\xi}_A+\iota \bar{\xi}_B V^{B}_{mA}=0.
\end{align}
The Killing vector near the North pole can be written as
\begin{align}
\upsilon^m\frac{\partial}{\partial x_m}=4c(x_1\frac{\partial}{\partial x_2}-x_2\frac{\partial}{\partial x_1})+4c(x_3\frac{\partial}{\partial x_4}-x_4\frac{\partial}{\partial x_3}).
\end{align}

Notice that near the South  pole our $N=2$ theory on squashed $S^4$ approaches topologically twisted theory with Omega deformation parameters $\epsilon_1=4c,\epsilon_2=4c$ \cite{Nekrasov:2003af,Nekrasov:2003rj},  and the contribution of these point-instantons is given by $Z_{inst}(a_0,\epsilon_1,\epsilon_2,\tau)$, where the parameter $\tau$ is defined by $\tau\equiv \frac{\theta}{2\pi}+\frac{4\pi \iota}{g_{YM}^2}$. 

Whereas near the North pole, the contribution of point anti-instantons is given by $Z_{inst}(a_0,\epsilon_1,\epsilon_2,\bar{\tau})$. Putting all together, the final form of the squashed $S^4$ partition function is
\begin{align}
Z=\int d \hat{a}_0 e^{-\frac{2\pi^3 \text{Tr}[a_0^2]}{c^2 g_{YM}^2}}|Z_{inst}|^2\prod_{\alpha\in \Delta_{+}}\Upsilon_{1}(\iota \hat{a}_0.\alpha)\Upsilon_{1}(-\iota \hat{a}_0.\alpha)\prod_{\rho\in R}\Upsilon_{1}(\iota \hat{a}_0.\rho+1)^{-1}.
\end{align}
\section{Conclusions}\label{conclusion}
 We have computed the partition function of $N=2$ SUSY on  squashed $S^4$ which admits $SU(2)\times U(1)$ isometry, using SUSY Localization technique. We find that  the full partition function is independent of the squashing parameters  as well as the other 
supergravity background fields.  
 
 The squashing functions independence of the one-loop part of the partition function,
which is obvious from the form of the relevant Killing vector $v$,  can perhaps be attributed
to the fact that in our squashed $S^4$  the theory is topologically twisted at the poles. This is because the $SU(2)_R$ symmetry which is generically broken down to $U(1)_R$ on the squashed $S^4$ excluding the poles, is again enhanced to $SU(2)_R$ at the poles. So this $SU(2)_R$ can be identified at the poles 
 with the $SU(2)$ Lorentz isometry to topologically twist the theory. The classical  part can be written as a total derivative and gives to a contribution which is independent of the squashing
parameters.
 
 It will be interesting to explain this independence along the same lines given
in \cite{Closset:2013vra}. That is to say, if we deform the vector multiplet and hypermultiplet actions around the round $S^4$ with respect to e.g. $f(r)$, it might be possible to write these deformed actions as  $Q$-exact terms separately. This $Q$-exactness of the deformed action will explain the independence of partition function of the parameter $f(r)$ in the sense of \cite{Closset:2013vra}. 
%This analogy is only suggestive because the authors in \cite{Closset:2013vra} discussed only
%the case of $N=1$ SUSY on compact four manifolds which are complex, whereas $S^4$ is not a complex manifold. 
However we have to consider perturbations around the round $S^4$,  unlike \cite{Closset:2013vra}, where it is perturbed around  flat $R^4$. 
 %We will report on this issue in future.

%It will also be interesting to use this more general squashing independence to study the properties of supersymmetric Renyi entropy on these four manifolds  as suggested in \cite{Crossley:2014oea,Huang:2014pda}.

\paragraph{Acknowledgements}
We would like to thank  G. Bonelli for discussions, and L.F. Alday
and L. Pando-Zayas for comments.
\bibliographystyle{elsarticle-num}
\nocite{*}
\bibliography{bibliography}

\end{document}